\documentclass[aps,twocolumn,showpacs,preprintnumbers,nofootinbib,prd,10pt,superscriptaddress]{revtex4-1}

\makeatletter
\def\l@subsubsection#1#2{}
\def\l@subsubsubsection#1#2{}
\makeatother

\usepackage{graphicx,amssymb,amsmath,amsthm,amsfonts,epsfig,epsf,fixmath}
\usepackage{comment}
\usepackage{mathrsfs}
\usepackage[usenames]{color}
\usepackage{epstopdf}
\usepackage{aas_macros}
\usepackage{bm}
\usepackage{dcolumn}
\usepackage{latexsym}
\usepackage{rotating}
\usepackage{longtable}

\usepackage{enumerate}
\usepackage{tensor,multirow}
\usepackage{url}
\usepackage[dvipsnames]{xcolor}
\usepackage[unicode]{hyperref}
\hypersetup{colorlinks=true, citecolor=MidnightBlue,
            linkcolor=MidnightBlue, urlcolor=MidnightBlue, linktocpage=true}

\newcommand{\dd}{\mathrm{d}}

\newcommand{\be}{\begin{equation}}
\newcommand{\eeq}{\end{equation}}
\newcommand{\ba}{\begin{align}}
\newcommand{\ea}{\end{align}}

\newcommand{\dder}[2]{\frac{\partial^2}{\partial #2^2} #1}

\newcommand{\GSSI}{Gran Sasso Science Institute (GSSI), I-67100 L’Aquila, Italy}
\newcommand{\GranSasso}{INFN, Laboratori Nazionali del Gran Sasso, I-67100 Assergi, Italy}

\begin{document}
\title{Probing time-dependent scalar wigs with extreme mass ratio inspirals}

\author{Matteo Della Rocca}
\email{matteo.dellarocca@phd.unipi.it}
\affiliation{Dipartimento di Fisica, Universit\`a di Pisa, Largo B. Pontecorvo 3, 56127 Pisa, Italy}
\affiliation{INFN, Sezione di Pisa, Largo B. Pontecorvo 3, 56127 Pisa, Italy}
\author{Georgios Antoniou}
\email{georgios.antoniou@uniroma1.infn.it}
\affiliation{Dipartimento di Fisica, Universit\`a di Pisa, Largo B. Pontecorvo 3, 56127 Pisa, Italy}
\affiliation{Sezione INFN Roma1, Roma 00185, Italy}
\affiliation{Dipartimento di Fisica, ``Sapienza'' Universit\`a di Roma, Piazzale Aldo Moro 5, 00185, Roma, Italy}
\author{Leonardo Gualtieri}
\email{leonardo.gualtieri@unipi.it}
\affiliation{Dipartimento di Fisica, Universit\`a di Pisa, Largo B. Pontecorvo 3, 56127 Pisa, Italy}
\affiliation{INFN, Sezione di Pisa, Largo B. Pontecorvo 3, 56127 Pisa, Italy}
\author{Andrea Maselli}
 \email{andrea.maselli@gssi.it}
\affiliation{\GSSI}
\affiliation{\GranSasso}

\begin{abstract} 
We investigate the gravitational wave emission from extreme mass ratio inspirals, key targets for the upcoming space-based detector LISA, considering the scenario where the lighter black hole in the binary is endowed by a long-lived, time-dependent scalar field configuration, known as a scalar wig. We develop a formalism to compute scalar perturbations for extreme mass ratio inspirals on circular orbits around Schwarzschild and Kerr black holes, and apply this framework to compute additional fluxes induced by the scalar wig as well as their dependence on the scalar field properties. Our computation provides strong indications that in this scenario the presence of the scalar field does not significantly affect the orbital motion and the gravitational waveform.
\end{abstract}

\maketitle

%--------------------------------------------------
\section{Introduction}\label{sec:introduction}
%--------------------------------------------------

Gravitational wave (GW) astronomy is a mature field of research. With over one hundred events observed to date, the merging of black holes (BHs) and neutron stars is now routinely detected by the LIGO/Virgo/KAGRA (LVK) collaboration \cite{LIGOScientific:2018mvr,LIGOScientific:2020ibl,LIGOScientific:2021usb,KAGRA:2021vkt}. This wide range of signals enables us to investigate binary coalescences in highly dynamic and strong-field gravitational regimes.

The sensitivity of current GW interferometers, however, restricts observations mostly to systems with nearly symmetric mass components, with the most asymmetric system detected so far having a mass ratio of $q\sim 1/10$ \cite{LIGOScientific:2020zkf,LIGOScientific:2020stg}.
The range of mass ratios detectable by GW detectors will significantly expand with the next generation of facilities, whether deployed on ground\,\cite{Punturo:2010zz,2015PhRvD..91h2001D}, in space as satellites\,\cite{2017arXiv170200786A,TianQin:2020hid}, or on the Moon as permanent observatories\,\cite{Jani:2020gnz,Ajith:2024mie}. Among the variety of science cases targeted by such detectors, fundamental physics studies stand out as primary goals\,\cite{Sathyaprakash:2019yqt,Kalogera:2021bya,Colpi:2024xhw,Barack:2018yly}.

The Laser Interferometer Space Antenna (LISA), in particular, will target signals peaking in the milli-Hz range, where binaries with significant mass asymmetry will evolve over timescales much longer than those expected for systems with comparable masses.

Among such sources, Extreme Mass-Ratio Inspirals (EMRIs), assembled by a stellar mass compact object (the secondary) orbiting around a massive companion BH (the primary), are expected to have mass ratios $q\sim 10^{-4}-10^{-6}$. EMRIs can stay in the LISA band for months to years, accumulating $\mathcal{O}(q^{-1})$ cycles on highly-relativistic trajectories before the plunge \cite{Babak:2017tow}. 
The rich harmonic content of the signal emitted allows for measurement of the source parameters with exquisite precision. This makes EMRIs valuable multi-purpose laboratories to probe fundamental physics\,\cite{Barack:2006pq,Sopuerta:2009iy,Babak:2017tow,Fransen:2022jtw,Raposo:2018xkf,Bena:2020see,Bianchi:2020bxa,Loutrel:2022ant,Piovano:2020ooe,Pani:2019cyc,Piovano:2022ojl,Datta:2019epe,Datta:2019euh,Maggio:2021uge,Pani:2010em,Macedo:2013qea,Destounis:2023khj,Datta:2019epe,Datta:2019euh,Liu:2020ghq,Burke:2020vvk,Rahman:2021eay,Maggio:2021uge,Miller:2024rca,Zi:2024dpi,Cardenas-Avendano:2024mqp,Zi:2023qfk,Chua:2018yng,Gair:2011ym,Cardoso:2011xi,Yunes:2011aa,Pani:2011xj,Canizares:2012is,Hannuksela:2018izj,Hannuksela:2019vip,Maselli:2020zgv,Barsanti:2022ana,Maselli:2021men,DellaRocca:2024pnm,Barsanti:2022vvl,Liang:2022gdk,Zhang:2023vok,Zi:2022hcc,Lestingi:2023ovn,Collodel:2021jwi,Brito:2023pyl,Duque:2023seg,Chen:2024ery,Fell:2023mtf,Barausse:2020rsu,Cardenas-Avendano:2024mqp}.

Recently, Ref.~\cite{Maselli:2020zgv} introduced a novel approach to explore EMRIs' potential in constraining the existence of new fundamental fields predicted by extensions of GR and of the Standard Model. This framework exploits the separation of scales provided by the EMRI mass asymmetry to simplify the description of their dynamics in theories involving scalar fields non-minimally coupled to gravity. As a result, Ref.\,\cite{Maselli:2020zgv} showed that deviations from the Kerr metric in the primary can be neglected at first order in the mass ratio $q$, and that changes in the binary evolution are uniquely controlled by the scalar charge of the secondary. 

This approach, which was recently incorporated into a rigorous self-force framework for computing waveform models accurate at post-adiabatic order\,\cite{Spiers:2023cip}, has been applied to study deviations in GW fluxes emitted by EMRIs in the presence of massless\,\cite{Barsanti:2022ana,DellaRocca:2024pnm} and massive scalar fields \cite{Barsanti:2022vvl}. 
Moreover, analyses based on Fisher matrix techniques \,\cite{Maselli:2021men,Tan:2024utr,Zhang:2022rfr,Guo:2022euk} and fully Bayesian methods \cite{Speri:2024qak} have explored LISA’s capability to constrain the scalar charge.

Studies conducted so far have focused on GR extensions. Within GR, no-hair theorems prohibit the existence of stationary scalar hair\,\cite{Bekenstein:1995un,Cardoso:2016ryw}. However BHs can support time-dependent, long-lived scalar field configurations, often referred to as \textit{scalar wigs} or \textit{gravitational atoms}\,\cite{Barranco:2011eyw,Barranco:2012qs,Baumann:2019eav}. These ultralight scalar configurations have gained significant interest, 
as potential dark matter candidates \cite{Hui:2019aqm,Clough:2019jpm,Cardoso:2022nzc}. 

Different formation mechanisms can lead to scalar wigs around BHs, from accretion of the scalar field in the galactic environment\,\cite{Hui:2019aqm,Clough:2019jpm,Cardoso:2022nzc,Budker:2023sex} to superradiance (see\,\cite{Brito:2015oca,Herdeiro:2015waa} and reference therein). 
Their length-scale is controlled by the mass of the scalar field. Indeed, a BH can support a scalar cloud as long as the Compton wavelength of the scalar field is larger than the horizon scale. We shall use geometric ($G=c=1$) units and denote the scalar field mass as $\mu_{\rm s}\hbar$; therefore, its Compton wavelength is $1/\mu_s$ and the condition for the formation of the scalar wig around a BH with mass ${\mathcal M}$ is ${\mathcal M}\mu_{\rm s}\lesssim 1$. 
We can express the mass of the scalar field in eV in terms of the dimensionless parameter $\mu{\cal M}$ as follows\,\cite{Brito:2015oca}:
\begin{equation}
\mu_{\rm s}[eV]\simeq \mu_{\rm s}{\mathcal M}\left(\frac{M_\odot}{{\mathcal M}}\right)\,10^{-10}{\rm eV}\,,
\label{eq:defev}
\end{equation}
with $M_\odot$ solar mass.

In recent years, EMRIs have attracted considerable attention as probes of scalar field clouds around the {\it primary} (e.g.\,\cite{Brito:2023pyl,Duque:2023seg}), with masses $M\sim (10^6-10^{10})M_\odot$. In this case, the scalar field mass can be as small as $\mu_{\rm s}\sim (10^{-19}-10^{-22})$ eV, probing the so-called {\it fuzzy dark matter} scenario\,\cite{Hu:2000ke}. 

In this article, we consider a different scenario, in which the scalar field has a larger mass, with a Compton wavelength of the order of - or a few orders of magnitude larger than - the scale of the  {\it secondary} body of an EMRI, with mass $m_{\rm p}\sim(10-100)M_\odot$.
Such a scalar field would not form a long-lived configuration around the primary BH, but could form it around the secondary, possibly affecting the EMRI waveform.

We shall consider scalar fields such that
\begin{equation}
    \frac{1}{M}\ll \mu_{\rm s}\lesssim\frac{1}{m_{\rm p}}\,,\label{eq:rangemu0}
\end{equation}
This corresponds, for $M\sim10^6\,M_\odot$ and $m_{\rm p}\sim10\,M_\odot$, to scalar field masses in the range $10^{-16}$ eV $\ll\mu_{\rm s}\lesssim\,10^{-11}$ eV. 
As we are going to discuss in Sec.\,\ref{sec:time_scales}, we shall focus on scalar field masses satisfying Eq.~\eqref{eq:rangemu0}, and such that 
the life-time of the scalar cloud is  large enough to use the adiabatic approximation - i.e. much larger than the orbital period.

In order to describe the EMRI, we extend the framework developed in\,\cite{Maselli:2020zgv}-\cite{Barsanti:2022vvl}. The main difference are that (i)  while in those papers the scalar field configuration was assumed to be stationary, in this case it is time-dependent; and (ii) at variance with\,\cite{Barsanti:2022vvl}, here the mass of the scalar field is much larger than $1/M$.

We shall show that this process can be described in terms of a rapidly oscillating scalar charge orbiting around the primary body, and that - despite the large frequency of the source - the scalar field emission is strongly suppressed. 
Our results provide strong indications that a scalar wig coupled with the secondary body does not significantly affect the EMRI orbital motion, and therefore its GW emission. 

We organize this work as follows. In Sec.~\ref{sec:theory} we present the theoretical framework we use to describe the secondary body endowed with a scalar wig. In Sec.~\ref{sec:adiabatic} we describe the EMRI inspiral, estimating  the energy loss due to the scalar field during the EMRI, in the case of a spherical symmetric scalar wig orbiting around a non-rotating primary BH. In Sec.~\ref{sec:disc} we  draw our conclusions. Finally, in App.~\ref{app:kerr} we generalise our derivation to the case of a rotating primary BH, and in App.~\ref{app:solvingwave} we discuss the solution of the wave equation in the large-frequency regime.

%--------------------------------------------------
\section{Theoretical Framework}\label{sec:theory}
%--------------------------------------------------

%
In this Section we introduce the theoretical framework to describe an EMRI in which the secondary body is endowed with a scalar wig, i.e. a long-lived configuration of an oscillating scalar field.
%
%-----------------------------------------------------------
\subsection{Action and field equations}\label{subsec:action}
%-----------------------------------------------------------
%
We consider a minimally coupled, massive, complex scalar field $\varphi$, described by the following action: 
\begin{equation}
 S\left[\textbf{g}, \varphi, \Psi \right] = S_0\left[\textbf{g}, \varphi\right] + S_{\rm m}\left[\textbf{g}, \varphi, \Psi\right]\ , \label{eq:action}
 \end{equation}
where   
\begin{equation}
    S_0 = \int d^4 x \frac{\sqrt{-g}}{16 \pi} \left(R - \frac{1}{2} \partial_\mu \varphi \partial^{\mu} \varphi^* - \frac{1}{2} \mu_{\rm s}^2 |\varphi|^2  \right)\ ,
 \end{equation}
$S_{\rm m}$ is the action of the matter fields $\Psi$, $R$ is the Ricci scalar, $\sqrt{-g}$ is the determinant of the metric tensor $\textbf{g}$, and $\mu_{\rm s}$ is the mass of the scalar field. 

The mass asymmetry in an EMRI naturally establishes a separation of scales between the primary mass, $M$, defining the external spacetime, and the secondary mass, $m_p$. The secondary can be effectively treated as a point particle, replacing $S_{\rm m}$ in Eq.~\eqref{eq:action} with the ``skeletonized action''\,\cite{eardley1975observable}  
\begin{equation}
S_p=-\int m(\varphi) \sqrt{g_{\alpha\beta}\frac{dy^{\alpha}_p}{d\lambda} \frac{dy^{\beta}_p}{d\lambda}}d\lambda\ ,
\end{equation}
where $m(\varphi)$ is a scalar function which depends on $\varphi$ at the location of the particle, and $y^\mu_{\rm p}(\lambda)$ is the particle's worldline, parametrized by the proper time $\lambda$.

Varying the action~\eqref{eq:action} with respect to the metric tensor and scalar field yields the field equations 
\begin{align}
& G_{\mu\nu} = 8 \pi T^{\rm scal}_{\mu \nu}+8 \pi T^{ p}_{\mu \nu}  \label{eq:fieldstens}\ ,\\
& ( \square - \mu_{\rm s}^2 ) \varphi =  16 \pi  \frac{\delta S_p}{\delta \varphi}\label{eq:fieldscal}\ ,
\end{align}
where $T^{\rm scal}_{\mu\nu}$ is the scalar-field stress-energy tensor and $T^{p}_{\mu\nu}$ is the stress-energy tensor for $S_p$. 
Eqns.~\eqref{eq:fieldstens}-\eqref{eq:fieldscal} can be solved perturbatively, using the mass ratio $q=m_p/M\ll1$ as an expansion parameter. 
At the leading order, the secondary acts as a perturbation of the background spacetime, and evolves on a sequence of geodesics until the plunge.

As discussed in the Introduction, we focus on ultra-light fields with $\mu_{\rm s}m_{\rm p}\lesssim1$, such that the secondary is endowed by a scalar cloud, formed through some physical process such as accretion or superradiance (even though in the case of superradiance, the scalar cloud is not spherical\,\cite{Brito:2015oca}, and our results should be considered as approximate). Conversely, since $\mu_{\rm s}M\gg 1$ the primary does not support any significant scalar field configuration\,\cite{Brito:2015oca}, and is then adequately described by the Kerr metric, with the background field fixed at a constant value $\varphi_0$.

The scalar field can then be expanded as $\varphi=\varphi_0+\varphi_1$, where $\varphi_1$ is the perturbation induced by the particle. At the leading order, $T^{\rm scal}_{\mu\nu}=O(\varphi_1^2)$ can be neglected and 
Eqs.~\eqref{eq:fieldstens}-\eqref{eq:fieldscal} reduce to:
\begin{align}
    & G^{\alpha\beta}=8\pi \int m(\varphi)\frac{\delta^{(4)}(x-x_{\rm p}(\lambda))}{\sqrt{-g}}\frac{\dd x_{\rm p}^\alpha}{\dd \lambda}\frac{\dd x_{\rm p}^\beta}{\dd \lambda} \dd \lambda \ ,\label{eq:E_general}\\
    & (\Box-\mu_s^2)\varphi=16\pi\int  m'(\varphi)\frac{\delta^{(4)}(x-x_{\rm p}(\lambda))}{\sqrt{-g}}\dd\lambda \,. \label{eq:KG_general}
\end{align}
The functions $m(\varphi)$, $m'(\varphi)$ have to  be evaluated at $\varphi=\varphi_0$. Indeed, while on the ``microscopic'' length-scale of the secondary body, $\varphi_0$ is the value of the scalar field in the buffer region around the particle, on the ``macroscopic'' length-scale of Eqs.~\eqref{eq:E_general}, \eqref{eq:KG_general}, $\varphi_0$ is the scalar field at the location of the particle.
We remark that on the length-scale of the secondary body, the tensor $T^{\rm scal}_{\mu\nu}$ contributes to the scalar wig around the secondary, and the mass $m_{\rm p}$ takes into account the contribution of the scalar cloud. We shall consider, however, scalar wigs in which the main contribution to the total mass comes from the BH.

%-----------------------------------------------------------
\subsection{Matching procedure for an oscillating scalar field}\label{sec:matching_procedure}
%-----------------------------------------------------------
%
In order to find the values of $m(\varphi_0)$ and $m'(\varphi_0)$, we match the solution of Eqs.~\eqref{eq:E_general}-\eqref{eq:KG_general} with the value of the metric and of the scalar field in a buffer region inside the world-tube of the secondary. This region has to be close enough to the body to be in the domain of its local frame $\{{\tilde x}^\mu\}$ and, at the same time, far enough to be outside its strong-field region:
\begin{equation}
    m_{\rm p}\ll \tilde r \ll r \ ,
\end{equation}
where $r\gtrsim M$ is the separation among the two bodies of the EMRI. In this frame 
\begin{align}
g_{\mu\nu}&=\eta_{\mu\nu}+\mathcal{O}\left[\left(\frac{m_{\rm p}}{\tilde r}\right)^2\right]+\mathcal{O}\left(\frac{\tilde r}{r}\right)\ .\label{eq:g_world_tube}
\end{align}

The scalar field forming a scalar wig around the secondary oscillates with a combination of proper frequencies $\{\omega_n\}_{n=0,\dots}$ \cite{Barranco:2012qs,Cardoso:2022nzc}.
In this case, the solution of the Klein-Gordon equation in the local frame reads:
\begin{align}
    \varphi({\tilde r})=\varphi_0&+\sum_{n} \frac{d_n m_p}{\tilde r}e^{i k_n \tilde r}e^{-i\omega_n \tilde t}\nonumber\\
    &+\mathcal{O}\left[\left(\frac{m_{\rm p}}{\tilde r}\right)^2\right]+\mathcal{O}\left(\frac{\tilde r}{r}\right)\label{eq:phi_world_tube} \,,
\end{align}
where $\tilde t={\tilde x}^0$ is the proper time of the particle, $k_n=\sqrt{\omega_n^2-\mu_{\rm s}^2}$, and $d_n$ is the scalar charge associated with the $n$-th mode.
Remarkably, for a scalar wig the real part of the frequency is always smaller than $\mu_s$, and the imaginary part is negative. Therefore, the factor $e^{ik_n{\tilde r}}$ in Eq.~\eqref{eq:phi_world_tube} decreases exponentially at large distances, and the cloud dissipates (by absorption from the BH) for $t\to\infty$.

Applying the matching procedure to Einstein's equations\,\eqref{eq:E_general} yields, as in the static case, $m(\varphi_0)=m_{\rm p}$. Moreover, from the Klein-Gordon's equation\,\eqref{eq:KG_general} expressed in the local frame we have
\begin{equation}
    \left(\tilde{\nabla}^2 - \frac{\partial^2}{\partial \tilde{t}^2}-\mu_{\rm s}^2\right)\varphi(\underline{\tilde x},\tilde{t})=- 4\pi \rho(\underline{\tilde x},\tilde{t}) \ ,
\label{eq:KleinGordon}\end{equation}
where 
\begin{equation}
        \rho(\underline{\tilde x},\tilde{t})=-4\int m'(\varphi)\frac{\delta^{(4)}(\underline{\tilde x}-\underline{\tilde x}_p(\lambda))}{\sqrt{-g}} \ \dd \lambda \,.
\end{equation}
As discussed above, the function $m'(\varphi)$ has to be evaluated at the particle's location, which yields:
\begin{equation}
\rho(\underline{\tilde x},\tilde{t})
  =-4m'(\varphi_0)\delta^{(3)}(\underline{\tilde x})\,.
\end{equation}
Then, matching the solution of Eq.~\eqref{eq:KleinGordon} with the expression of the scalar field in the local frame, Eq.~\eqref{eq:phi_world_tube}, we obtain
\begin{equation}
\frac{m'(\varphi_0)}{m_{\rm p}}=-\frac{1}{4}\sum_n d_n e^{-i\omega_n \tilde t}
\,.\label{eq:mp}
\end{equation}
Note that for time-independent charges studied in \cite{Maselli:2020zgv}-\cite{Barsanti:2022vvl}, the matching procedure leads to $m'(\varphi_0)/m_{\rm p}=-d/4$, for both massless and massive scalars, with $d$ being the secondary constant charge.
In the case of scalar wigs, the latter is replaced by a time varying charge, which oscillates with the proper frequencies of the wig.

Finally, by replacing Eq.~\eqref{eq:mp} into 
Eqns.~\eqref{eq:E_general}-\eqref{eq:KG_general} 
we find
\begin{equation}
    G^{\alpha\beta}=8\pi m_p\int \frac{\delta^{(4)}(x-x_{\rm p}(\lambda))}{\sqrt{-g}}\frac{\dd x_{\rm p}^\alpha}{\dd \lambda}\frac{\dd x_{\rm p}^\beta}{\dd \lambda} \dd \lambda \,,\label{eq:E1}
\end{equation}
which describe the standard linear order perturbations 
sourced by the EMRI secondary on the Kerr background, 
and
\begin{equation}
    (\Box-\mu^2_{\rm s})\varphi=T\ ,
    \label{eq:KG1}
\end{equation}
with 
\begin{equation}
    T=-\frac{4\pi m_p}{\sqrt{-g} {\dot t}_{\rm p}}    \delta^{(3)}(\underline x-\underline{x}_{\rm p}(t))\sum_{n}d_n e^{-i\omega_n \tilde{t}\left(\underline{x}_{\rm p}\right)}
    \,.\label{eq:defT1}
\end{equation}
Here $\{x^\mu\}=(t,r,\theta,\phi)$ correspond to the Boyer-Lindquist frame of the exterior Kerr metric\,\eqref{eq:Kerr:metric}, $x_{\rm p}^\mu$ are the coordinates of the particle, and we denote with a dot derivatives with respect to its proper time $\tilde t$.

For the sake of simplicity, we shall consider the case in which only the fundamental ($n=0$) mode is excited, such that the oscillation is monochromatic, with frequency $\omega_0=\omega_0^R-i/\tau$, where $\tau=-1/\omega_0^I>0$ is the damping time of the configuration. Therefore,
\begin{equation}
    T=-\frac{4\pi m_p}{\sqrt{-g} {\dot t}_{\rm p}}    \delta^{(3)}(\underline x-\underline{x}_{\rm p}(t))d_0 e^{-i\omega_0^R \tilde{t}{(\underline{x}_{\rm p})}}e^{-\tilde{t}_{\rm p}/\tau}
    \label{eq:defT2}
\end{equation}
where $\tilde{t}_{\rm p}=\tilde{t}(x_{\rm p})$.

%--------------------------------------------------
\section{Adiabatic inspiral}\label{sec:adiabatic}
%--------------------------------------------------
We solve Eqs.~\eqref{eq:E1}-\eqref{eq:KG1} in the adiabatic limit, where the timescale of the orbital parameter variation is longer than the orbital period. Under this approach, the motion of the secondary can be approximated as a sequence of geodesics around the primary, in which the transitions  between geodesics are governed by the energy and angular momentum fluxes. 

The primary is described by the Kerr metric,
\begin{align}
    &ds^2=-\left(1-\frac{2Mr}{\Sigma}\right)dt^2-\frac{4Mra\sin^2\theta}{\Sigma}dt d\phi+\Sigma d\theta^2 +
    \nonumber\\
    &+\frac{\Sigma}{\Delta}dr^2+\left(r^2+a^2+\frac{2Mra^2}{\Sigma}\sin^2\theta\right)\sin^2\theta d\phi^2 \ ,
\label{eq:Kerr:metric}
\end{align}
where $\Delta=r^2+a^2- 2M r$, $\Sigma=r^2+a^2\cos^2\theta$.

For simplicity, here we consider circular, equatorial geodesics, although our framework can be readily extended to accommodate more complex orbital configurations.
In this case, the secondary's location on a geodesic of radius $r_0$ is given by 
\begin{equation}
    r_{\rm p}(t)=r_0\ , \quad \phi_{\rm p}(t)=\Omega_\phi t\ ,\quad \theta_{\rm p}(t)=\frac{\pi}{2}\label{eq:circ_eq_motion} \ ,
\end{equation}
where $\Omega_\phi$ depends on the mass $M$ and the angular momentum $J$ of the primary BH, and is given by 
\begin{equation}
    \Omega_\phi
    =\epsilon\frac{\sqrt{M}}{r_0^{3/2}+\epsilon a \sqrt{M}}\ ,\label{eq:Kerr:circeqfreq}
\end{equation}
with $a=J/M$, and $\epsilon=\pm1$ for prograde/retrograde orbits. 
Finally, it is useful to introduce the Lorentz factor $\gamma=t/\tilde{t}_{\rm p}$:
\begin{equation}
\gamma=\big|g_{tt}+g_{t\phi}\Omega_\phi+g_{\phi\phi}\Omega_\phi^2\big|^{-1/2}\label{eq:tgamma}\,.
\end{equation}
For EMRIs emitting in the LISA band, $r_0\sim 10 M$, which leads to $\gamma\sim 1$. However, as we are going to show, it gives a very important contribution to the perturbation equations.

If the primary BH is non-rotating, the background spacetime is the Schwarzschild metric
\begin{equation}
ds^2=-f(r)dt^2+f^{-1}(r)dr^2+r^2(d\theta^2+\sin^2\theta d\phi^2)\,,
\end{equation}
with $f=1-2M/r$, and the Lorentz factor is $\gamma=\left(1-{3M}/{r_0}\right)^{-1/2}$.
%
%--------------------------------------------------
\subsection{Time-scale approximations}\label{sec:time_scales}
%--------------------------------------------------
%
The evolution of the scalar wig during the EMRI inspiral 
is characterized by three timescales:
\begin{itemize}
\item[(i)] The period of oscillation of the scalar field, $\Upsilon=2\pi\gamma/\omega_0^R\simeq2\pi/\mu_{\rm s}$.
\item[(ii)] The life-time of the scalar wig, $\tau$. For $l=0$, analytical estimates of the imaginary part of the bound state frequency give $m_{\rm p}\omega_0^I\sim 8(\mu_{\rm s} m_{\rm p})^6$ \cite{Pani:2012bp,Barranco:2012qs,Cardoso:2022nzc}, while for $l>0$ the timescale is larger. Therefore, for the most efficiently damped modes ($\ell=0$),
    \begin{equation}
       \tau\sim 0.1(\mu_{\rm s} m_{\rm p})^{-6}m_{\rm p}\,.\label{eq:extbart}
    \end{equation}
For simplicity, we shall only consider an $\ell=0$ mode, i.e. a {\it spherically symmetric} scalar wig. Note also that the back-reaction of the scalar cloud can affect $\omega_0$, therefore the frequencies and damping times of quasi-bound states should be considered as approximate estimates of the correct values.
\item[(iii)] The orbital period $P=\frac{2\pi}{\Omega_\phi}\sim 100M$ (see Eq.\,\eqref{eq:Kerr:circeqfreq}).
\end{itemize}
Comparing the timescales we have:
\begin{align}
    \frac{\tau}{P}&\sim10^{-3}\,(\mu_{\rm s} m_{\rm p})^{-6}q\,,\\
    \frac{\Upsilon}{P}&\sim6\times 10^{-2}\frac{q}{m_{\rm p}\mu_{\rm s}}\,.
\end{align}
Hereafter we consider a typical EMRI with $M=10^6\,M_\odot$, $m_{\rm p}=10\,M_\odot$, i.e. $q=10^{-5}$. 
We also focus on scalar field masses consistent with Eq.~\eqref{eq:rangemu0}, such that $\mu_{\rm s}M\ge 100$ and $\mu_{\rm s}m_{\rm p}\lesssim 1$.
Moreover, working in the adiabatic approximation requires $P/\tau\ll 1$. These conditions are satisfied by choosing $0.001\le \mu_{\rm s}m_{\rm p}\le 0.02$, corresponding to $10^{-14}{\rm eV}\le \mu_{\rm s}\le 2\times 10^{-13}{\rm eV}$.

In this regime we always have $\Upsilon\ll P$, i.e. the scalar field oscillates much more rapidly than the orbital frequency. Moreover, since $P\ll\tau$ we can neglect the decay of the scalar field cloud during the orbital evolution, and compute the corresponding flux ignoring the exponential factor, $e^{-\tilde t/\tau}$, in the source term of Eq.~\eqref{eq:KG1}. 

It is worth noting that the real part of the eigen-frequencies of the scalar wig, $\omega_n^R$, is always smaller than 
the scalar field mass.
However, for the values of $\mu_{\rm s}m_{\rm p}$ we consider, these these quantities are very close to
each other and \cite{Dolan:2007mj,Barranco:2017aes,Cardoso:2022nzc}
\begin{equation}
\omega_0^R\simeq\mu_{\rm s}\left(1-\frac{1}{2}(\mu_{\rm s}m_{\rm p})^2\right)\,.\label{eq:omegamu}
\end{equation}
%
%--------------------------------------------------
\subsection{Perturbation equations with source}
\label{subsec:pert}
%--------------------------------------------------
%
In order to compute the energy and angular momentum fluxes driving the EMRI evolution, we solve the equations for the metric and the scalar perturbations.

Since the equations for the gravitational sector, Eqs.\,\eqref{eq:E1}, are not affected by the scalar field and have been extensively studied in the literature, we do not discuss them here (see e.g. App. B of\,\cite{Barsanti:2022ana} and references therein). 
We focus on the solution of the scalar field equation, Eq.\,\eqref{eq:KG1}, with the source\,\eqref{eq:defT2} which, with Eq.~\eqref{eq:tgamma}, can be written as
\begin{equation}
    T=-\frac{4\pi m_p}{\sqrt{-g} {\dot t}_{\rm p}} \delta^{(3)}(\underline x-\underline{x}_p(t))d_0
    e^{-i\omega_0^R t/\gamma}
    \ .\label{eq:defT}
\end{equation}
For sake of simplicity, in the following we focus on perturbations around a Schwarzschild background, referring the reader to App.\,\ref{app:kerr} for the full extension to the Kerr case.
Expanding the scalar field $\varphi$ and the source term $T$ in spherical harmonics,
\begin{align}
    \varphi(t,r,\theta,\phi)&=\sum_{\ell m}\psi_{\ell m}(t,r)Y^{\ell m}(\theta,\phi) \, ,\label{eq:harmphi}\\
    T(t,r,\theta,\phi)&=\sum_{\ell m}\mathcal{T}_{\ell m}(t,r)Y^{\ell m}(\theta,\phi) \,,\label{eq:harmT}
\end{align}
the angular dependence decouples, and the Klein-Gordon equation\,\eqref{eq:KG1} reduces to
\begin{equation}
    \left[\dder{}{r_*}-\dder{}{t}-V_l(r)\right]R_{\ell m}=r\left(1-\frac{2M}{r}\right)\mathcal{T}_{\ell m}\label{eq:eqRT}
\end{equation}
where we have defined the auxiliary function $R_{\ell m}=r\psi_{\ell m}$, $r_*=r+2M \log\left|\frac{r}{2M}-1\right|$, and
\begin{equation}
    V_{l}=\left(1-\frac{2M}{r}\right)\left(\frac{l (l+1)}{r^2}+\frac{2M}{r^3}+\mu^2\right) \,.
\label{eq:V_ReggWheeler}
\end{equation}
We also introduce the Fourier transforms
\begin{align}
    R_{\ell m}(t,r)&=\int_{-\infty}^{+\infty}\dd \omega\  \tilde{R}_{\ell m\omega}(r)e^{-i\omega t} \, , \label{eq:fourierR}\\
    rf\mathcal{T}_{\ell m}(t,r)&=\int_{-\infty}^{+\infty}\dd \omega\  \tilde{\tau}_{\ell m\omega}(r)e^{-i\omega t} \,,\label{eq:fourierT}
\end{align}
such that Eq.~\eqref{eq:eqRT} becomes
\begin{equation}
   \left[\dder{}{r_*}+\omega^2-V_l(r)\right]\tilde{R}_{\ell m\omega}=\tilde{\tau}_{\ell m\omega} \,.
\label{eq:Einstein_Klein_Gordon_Fourier}
\end{equation}
The solution of Eq.~\eqref{eq:Einstein_Klein_Gordon_Fourier} can be found using the Green function approach (for brevity, we neglect the indexes $\ell m\omega$): 
\begin{equation}
    \tilde{R}=Z^+(r_*)\tilde{R}^+(r_*)+Z^-(r_*)\tilde{R}^-(r_*)\ ,
    \label{eq:solRt}
\end{equation}
where 
\begin{align}
        Z^+(r_*)&=\frac{1}{W}\int_{-\infty}^{r_*}\dd r_*' \ \tilde{\tau}(r_*')\tilde{R}^-(r_*')\ ,\label{eq:Zplus_S}\\
        Z^-(r_*)&=\frac{1}{W}\int_{r_*}^{+\infty}\dd r_*'\  \tilde{\tau}(r_*')\tilde{R}^+(r_*')\,. \label{eq:Zminus_S}
\end{align}
$\tilde{R}^\pm$ are two independent solutions of the homogeneous problem asociated to Eq.~\eqref{eq:Einstein_Klein_Gordon_Fourier}, such that $\tilde{R}^+$ ($\tilde{R}^-$) is a purely outgoing (ingoing) wave solution at spatial infinity (horizon). 
The Wronskian $W=\tilde{R}^+ \tilde{R}^-_{,r_*}-\tilde{R}^-\tilde{R}^+_{,r_*}$ is constant.  

The asymptotic behaviour of  the functions ${\tilde R}_{\ell m}^\pm$ is as follows.
For $r_*\to-\infty$,
\begin{align}
    \tilde{R}^+&\to B_{out}^+ e^{i\omega r_*} + B_{in}^+ e^{-i\omega r_*}\nonumber\\
    \tilde{R}^-&\to e^{-i\omega r_*}\,,\label{eq:asymptoci_H}
\end{align}
while for $r_*\to+\infty$,
\begin{align}
    \tilde{R}^+&\to e^{ik_\omega r_*}\nonumber\\
    \tilde{R}^-&\to A_{out}^- e^{ik_\omega r_*}+A^-_{in}e^{-ik_\omega r_*}\,, \label{eq:asymptoci_inf0}
\end{align}
where $k_\omega=\sqrt{\omega^2-\mu_{\rm s}^2}$. With this choice, $W=-2ik_\omega A^-_{in}$.

If $\omega<\mu_{\rm s}$ (like in our case, as we shall discuss below), $k_\omega$ is purely imaginary; in this case, we define $k_\omega=i\kappa_\omega$ where $\kappa_\omega=\sqrt{\mu_{\rm s}^2-\omega^2}$ is real and positive. Therefore, for $\omega<\mu_{\rm s}$ the behaviour of $\tilde R$ for $r_*\to+\infty$ is
\begin{align}
    \tilde{R}^+&\to e^{-\kappa_\omega r_*}\nonumber\\
    \tilde{R}^-&\to A_{out}^- e^{-\kappa_\omega r_*}+A^-_{in}e^{\kappa_\omega r_*}\,,\label{eq:asymptoci_inf}
\end{align}
and $W=2\kappa_\omega A^-_{in}$. Note that the choice $k_\omega=-i\kappa_\omega$ would have been unphysical, since it would lead to a function $\tilde R(r)$ in Eq.~\eqref{eq:solRt} asymptotically divergent.

The source $\tilde\tau$ in Eqs.~\eqref{eq:Zplus_S}-\eqref{eq:Zminus_S} 
is given by (restoring the indexes $\ell m\omega$):
\begin{align}
\tilde{\tau}_{\ell m\omega}&=-\frac{4\pi {d}_0 m_{\rm p}}{r_0}Y_{\ell m}(\pi/2,0)\left(1-\frac{2M}{r_0}\right)\sqrt{1-\frac{3M}{r_0}}\nonumber\\
&\times\delta(r-r_0)\delta(\omega-\bar\omega_m)\ ,
\label{eq:hat_tau_A}
\end{align}
where $\bar\omega_m=\omega^R_0/\gamma+m \Omega_\phi$.

We remark that the orbital frequency $\Omega_\phi\sim O(1/M)$ is much smaller than the oscillation frequency of the scalar field $\omega_0^R\sim O(1/m_{\rm p})$. Thus, the characteristic frequency of the source, $\bar\omega_m$, is dominated by the latter. A simple computation shows that (despite the fact that $\omega_0^R$ is close to $\mu_{\rm s}$ (see Eq.\,\eqref{eq:omegamu}), $\bar\omega_m<\mu_{\rm s}$ for $m\lesssim 400$. Since the emission for very large harmonic indexes is strongly suppressed (see e.g.\,\cite{Fujita:2005kng,Drasco:2005kz}), in the following we assume that $\bar\omega_m<\mu_{\rm s}$. 

By replacing the source term in Eqs.~\eqref{eq:Zplus_S}-\eqref{eq:Zminus_S}, using $\dd r_*/\dd r=1/f(r)$, we get
\begin{equation}
Z_{\ell m\omega}^{\pm}(r)=\frac{1}{r_0}{\mathcal C}_{\ell m\omega}{\tilde R}^\mp_{\ell m\omega}(r_{0*})\Theta(\pm(r-r_0))\delta(\omega-\bar\omega_m)\label{eq_exprZ}
\end{equation}
where $r_{0*}=r_*(r_0)$ and
\begin{equation}
    {\mathcal C}_{\ell m\omega}=-\frac{4\pi d_0 m_{\rm p}}{W}\sqrt{1-\frac{3M}{r_0}} Y_{\ell m}(\pi/2)
\,. \label{calCa}
\end{equation}
From Eqs.\,\eqref{eq:solRt}, \eqref{eq:asymptoci_H} and \eqref{eq:asymptoci_inf}, we find that for $r\to2M$
\begin{equation}
\tilde R_{\ell m\omega}(r)\sim \frac{e^{-i\omega r_*}}{r_0}{\mathcal C}_{\ell m\omega}{\tilde R}^+_{\ell m\omega}(r_{0*})\delta(\omega-\bar\omega_m)\label{asRH}
\end{equation}
while for $r\to\infty$, given $\bar\omega_m<\mu_{\rm s}$,
\begin{equation}
\tilde R_{\ell m\omega}(r)\sim \frac{e^{-\kappa_\omega r_*}}{r_0}{\mathcal C}_{\ell m\omega}{\tilde R}^-_{\ell m\omega}(r_{0*})\delta(\omega-\bar\omega_m)\,.\label{asRinf}
\end{equation}
%
%-------------------------------------------------
\subsection{Energy flux}\label{subsec:flux}
%-------------------------------------------------

%
The scalar field energy fluxes at infinity and at the horizon are given by (see e.g.\,\cite{Warburton:2010eq}):
\begin{equation}
\dot E_\pm=\frac{\dd E_\pm}{\dd t}=\mp \lim_{r_*\to\pm\infty}\int \dd \Omega\  r^2 T_{tr}g^{rr}
\label{eq:E_dot_SC}
\end{equation}
where for a Schwarzschild background (see App.\,\ref{app:kerr} for the extension to a Kerr background) $r^2g^{rr}=r^2f(r)$ and 
\begin{equation}
    T_{tr}=\frac{1}{16\pi}\left(\partial_t \varphi\partial_r \varphi^*+\partial_t \varphi^*\partial_r \varphi\right) \,.
\label{eq:Trt_SC}
\end{equation}
Replacing equations \eqref{eq:harmphi} and \eqref{eq:fourierR} into Eq.~\eqref{eq:E_dot_SC} we obtain
\begin{align}
\dot{E}_{\pm}=&\frac{1}{16\pi}\lim_{r_*\to\pm\infty}
\sum_{\ell m}\int_{-\infty}^{+\infty}\dd\omega
\int_{-\infty}^{+\infty}\dd\omega'{\tilde R}_{\ell m\omega}(r)
\nonumber\\
&\times {\tilde R}^*_{\ell m\omega'}(r)_{,r_*}e^{-i(\omega-\omega')t}(-i\omega)+c.c.\,.
\label{eq:solEdot}
\end{align}
Finally, using the solutions \eqref{asRH}-\eqref{asRinf} we find for the flux at infinity 
\begin{align}
\dot E_+
\simeq\frac{1}{16\pi r_0^2}&\lim_{r_*\to+\infty}\sum_{\ell m}
i\bar\omega_m\kappa_{\bar\omega_m}|\mathcal{C}_{\ell m}(\bar\omega_m)|^2\times\nonumber\\
&\times|\tilde{R}^-_{\ell m\bar\omega_m}(r_0)|^2
e^{-2\kappa_{\bar\omega_m}r_*}+c.c.\simeq 0\,,\label{eq:Ef_dotfin}
\end{align}
where we have used the fact that 
$\kappa_{\bar\omega_m}$ is real and positive since $\bar\omega_m<\mu_s$. The flux at the horizon 
of the primary BH is given by:
\begin{align}
&{\dot E_-}\simeq
\sum_{\ell m}\frac{{\bar\omega_m}^2}{8\pi r_0^2}
|\mathcal{C}_{\ell m}(\bar\omega_m)|^2|\tilde{R}^+_{\ell m\bar\omega_m}(r_0)|^2\ , \label{eq:Em_dotfin}
\end{align}
and by replacing Eq.\,\eqref{calCa}:
\begin{align}
    {\dot E_-}=&\frac{2d_0^2 m_p^2\pi}{r_0^2}\left(1-\frac{3M}{r_0}\right) \sum_{\ell m} \bar{\omega}^2_m {\mathcal R}_{\ell m}(r_0)
    Y^2_{\ell m}\big(\pi/2\big) \,,\label{eq:Em_dot}
\end{align}
where we have defined
\begin{equation}
{\mathcal R}_{\ell m}(r_0)=
\frac{|
\tilde R_{\ell m\bar{\omega}_m}^+(r_0)|^2}{|W|^2}\,.
\label{eq:ratio_R/W}
\end{equation}
Equations \eqref{eq:Ef_dotfin} and \eqref{eq:Em_dotfin}-\eqref{eq:Em_dot} are among the main results of this paper, and show that the computation of the scalar energy flux for an EMRI in which the secondary is endowed by a scalar wig reduces to determining ${\mathcal R}_{\ell m}(r_0)$. That is finding the solution of the homogeneous equation of Eq.~\eqref{eq:Einstein_Klein_Gordon_Fourier}, with boundary conditions~\eqref{eq:asymptoci_H},~\eqref{eq:asymptoci_inf}, evaluated at $r=r_0$.

The numerical integration of such equation is challenging due to the different scales involved: $\omega^{-1}$, $\mu_{\rm s}^{-1}\ll M$, while $r_0\sim (6-15)\,M$ for a source in the LISA band.
In this regime, the terms $\,l(l+1)/r^2$, $M/r^3$ in the scattering potential $V_l(r)$ can be neglected (unless $l\sim O(100)$ or larger, in which case the emission is strongly suppressed, see e.g.\,\cite{Fujita:2005kng,Drasco:2005kz}), and 
\begin{equation}
    \left[\frac{d^2}{dr_*^2}+{\mathcal V}_{\omega}(r)\right]{\tilde R}^{\pm}_{\ell m\omega}=0\label{eq:homlargew}
\end{equation}
with the effective potential
\begin{equation}
%V_l(r)\simeq
    {\mathcal V}_{\omega}(r)=\omega^2-V_l(r)\simeq\omega^2-\left(1-\frac{2M}{r}\right)\mu_{\rm s}^2\,.\label{eq:calV}
\end{equation}
An analysis of Eq.~\eqref{eq:Einstein_Klein_Gordon_Fourier} in this case 
(see Appendix\,\ref{app:solvingwave}) shows that ${\mathcal R}_{\ell m}(r_0)$ 
and hence the scalar flux at the horizon 
critically depends on the sign of the effective potential\,\eqref{eq:calV} at 
$r=r_0$. In particular, $\dot E_-$ can be large if ${\mathcal V}(r_0)>0$, while 
it is negligible if ${\mathcal V}(r_0)<0$. 

We remind that $r_0\sim(6-15)M$ during the EMRI 
evolution, while $\mu_s,\omega\gg M^{-1}$, and $\omega$ is sligthly smaller than $\mu$. Therefore, ${\mathcal V}_\omega(r_0)$ is negative for large values of $r_0$ but it may become positive for $r_0<\bar r$ given by ${\mathcal V}_\omega(\bar r)=0$. In this scenario, the scalar emission is ``activated'' when the inspiralling body crosses the threshold radius $\bar r$.

However, a closer inspection of the scalar field equation shows that this is not the case. Indeed, since $\omega={\bar\omega}_m=\omega_0^R/\gamma+m\Omega\phi$, with $m\Omega_\phi$ negligible (see Appendix\,\ref{app:solvingwave}) and $\gamma=(1-3M/r_0)^{-1/2}$, leading to a correction of the frequency which dominates over the mass redshift  at $r=r_0$:
\begin{equation}
{\mathcal V}_{\bar\omega_m}(r)\simeq\left(1-\frac{3M}{r_0}\right)(\omega_0^{R})^2-\left(1-\frac{2M}{r}\right)\mu^2\,,\label{eq:calV1}
\end{equation}
and ${\mathcal V}(r_0)<0$ for all values of $r_0$. Therefore, the scalar emission is always suppressed.
As shown in Appendix~\ref{app:solvingwave}, calculations performed in this section can be straightforwardly generalised to the case of Kerr background. 

Moreover, in App.\,\ref{app:proof_neg} we show that the effective potential in the perturbation equation, evaluated at the location of the particle, is negative for a general orbit (in Kerr background), for the range of masses considered in this paper. Although - strictly speaking - we do not have explicitly computed the energy flux in this case,  our results suggest that even for EMRIs on general orbits, the  emission of a scalar wig coupled to the secondary body is strongly suppressed.
%%%%%%%%%%%%%%%%%%%%%%%%%%%%%%%%%%%%%%%%%%%%%%%%%%%%%%%%%
\section{Conclusions}\label{sec:disc}
%%%%%%%%%%%%%%%%%%%%%%%%%%%%%%%%%%%%%%%%%%%%%%%%%%%%%%%%%
In this article we have studied how the presence of a scalar wig around the secondary body of an EMRI could affect its orbital evolution. We have considered a scalar field with a Compton wavelength much smaller than the length-scale of the primary body, and few orders of magnitude larger of that of the secondary; the further requirement of working in the 
adiabatic regime leads to $0.001\le \mu_{\rm s}m_{\rm p}\le 0.02$, i.e. $10^{-14}{\rm eV}\le \mu_{\rm s}\le 2\times 10^{-13}{\rm eV}$.

Since $\mu_{\rm s}M\gg 1$, a {\it stationary} scalar cloud of such large mass would obviously have negligible emission, since in this 
case the emission channel would be  dipolar emission, driven by the orbital motion at frequencies $\Omega_\phi$\,\cite{Barsanti:2022vvl}, which in this case are always much smaller than $\mu_{\rm s}$.  The case of a scalar wig, which is an {\it oscillating} cloud, is more subtle, since in this case the main emission channel would be  the monopolar oscillation of the wig, at frequency $\omega_R^0\sim \mu_{\rm s}$.

However, we have found that - in the range of scalar field masses considered - the effect of the scalar wig in the EMRI evolution is negligible as well. 
Indeed, even though the rapidly oscillating charge may in principle lead to a large scalar emission, this would require the oscillation frequency at the location of the secondary to be larger than the redshifted scalar mass, and we find that this does not occur during the EMRI evolution.
Strictly speaking, we have proven this result only for a quasi-circular, equatorial EMRI on a Kerr background, although we have strong indications that the same holds for more general orbital configurations.

We stress that the suppression of scalar emission is due to the separation of scales. In the case of comparable-mass binaries (or possibly of intermediate mass-ratio inspirals), we do not expect the scalar emission to be suppressed. In this case, which deserves further investigation, the emission could be due to a combination of different mechanisms: the dipolar emission driven by orbital motion, and the monopolar emission due to the scalar charge oscillation. These phenomena could lead both to a modification of the orbital motion, and to the depletion of the scalar cloud itself.

%%%%%%%%%%%%%%%%%%%%%%%%%%%%%%%%%%%%%%%%%%%%%%
\acknowledgments
We thank F. Duque, T. Sotiriou, V. Cardoso, P. Pani, M. Vaglio, R. Brito, S. Barsanti, and N. Warburton for useful comments and discussions.
We acknowledge financial support from the EU Horizon2020 Research and Innovation Programme under the MarieSklodowska-Curie Grant Agreement No. 101007855.
G. A. acknowledges support by the INFN TEONGRAV initiative.
A   M. acknowledges support from the ITA-USA Scienceand Technology Cooperation program, supported by theMinistry of Foreign Affairs of Italy (MAECI). 
A. M.acknowledges financial support from MUR PRIN Grants No. 2022-Z9X4XS and No. 2020KB33TP.
%%%%%%%%%%%%%%%%%%%%%%%%%%%%%%%%%%%%%%%%%%%%%%
\appendix
%%%%%%%%%%%%%%%%%%%%%%%%%%%%%%%%%%%%%%%%%%%%%%%%%%%%%%%%%
\section{Kerr spacetime}\label{app:kerr}
%%%%%%%%%%%%%%%%%%%%%%%%%%%%%%%%%%%%%%%%%%%%%%%%%%%%%%%%%
In this Appendix we extend the derivation of Secs.\,\ref{subsec:pert}, \ref{subsec:flux} to the case of a Kerr background.

The Klein-Gordon equation\,\eqref{eq:KG1} on Kerr background coincides with the Teukolsky equation for massive scalar fields\,\cite{1973ApJ...185..635T},
\begin{align}
&\left[ \frac{(r^2 + a^2)^2}{\Delta} - a^2 \sin^2 \theta \right] \frac{\partial^2 \varphi}{\partial t^2}  + \frac{4 M a r}{\Delta} \frac{\partial^2 \varphi}{\partial t \partial \phi}\nonumber\\
&+ \left[ \frac{a^2}{\Delta} - \frac{1}{\sin^2 \theta} \right] \frac{\partial^2 \varphi}{\partial \phi^2} - \frac{1}{\sin \theta} \frac{\partial}{\partial \theta} \left( \sin \theta \frac{\partial \varphi}{\partial \theta} \right)\nonumber\\
&  \frac{\partial}{\partial r} \left( \Delta \frac{\partial \varphi}{\partial r} \right) +\mu^2 \Sigma \varphi= -\Sigma T\,.
\label{eq:Teukolsky_master_mquation}
\end{align}
Expanding the scalar field and the source in  spheroidal harmonics,
\begin{align}
    \varphi(t,r,\theta,\phi)&=\sum_{\ell m}\psi_{\ell m}(t,r)S^{\ell m}(\theta,\phi)\ ,\label{eq:kerr_harm_phi}\\
    \Sigma(r,\theta)T(t,r,\theta,\phi)&=\sum_{\ell m}\mathcal{T}_{\ell m}(t,r)S^{\ell m}(\theta,\phi)\label{eq:kerr_harm_T} \,,
\end{align}
defining ${R}_{l m}(t,r)=\rho(r) \psi_{l m}(t,r)$ with $\rho(r)\equiv \sqrt{r^2+a^2}$,
and introducing the Fourier transforms
\begin{align}
    \tilde R_{\ell m}(t,r)=
    &\int \dd \omega\, \tilde{R}_{\ell m\omega}(r) e^{-i\omega t},\label{eq:kerr_four_R}\\
    \frac{\Delta(r)}{\rho(r)^{3}}\mathcal{T}_{\ell m}(t,r)=
    &\int \dd \omega\, \tilde{\tau}_{\ell m\omega}(r) e^{-i\omega t}\,,\label{eq:kerr_four_T}
\end{align}
we find
\begin{equation}
    \left[\dder{}{r_*}+\mathcal{V}_{\ell m\omega}(r)\right]\tilde{R}_{\ell m\omega}=\tilde{\tau}_{\ell m\omega} \, ,
    \label{eq:X}
\end{equation}
where $r^\star$ is the tortoise coordinate $dr^\star/dr=\rho^2/\Delta$. The potential $\mathcal{V}_{\ell m\omega}(r)$ is given by\,\cite{Barsanti:2022vvl}
\begin{align}
    \mathcal{V}_{\ell m\omega}(r)=&\left(\omega-\frac{am}{\rho^2}\right)^2-\frac{\Delta}{\rho^4}\left(\lambda_{\ell m}+\frac{2Mr^3}{\rho^4}\right.\nonumber\\
    &\left.+2ma(\kappa_\omega-\omega)+\frac{a^2}{\rho^2}+\mu_{\rm s}^2\rho^2\right)\,, 
\label{eq:potKerr}
\end{align}
where $\lambda_{\ell m}$ is the eigenvalue of the spin-weighted spheroidal harmonic $S_{\ell m\omega}$.
The general solution of this equation can then be written as 
\begin{equation}
 \tilde{R}_{\ell m\omega}=Z^{-}_{\ell m\omega}(r_*)\tilde{R}^{-}_{\ell m\omega}(r_*)+Z^{+}_{\ell m\omega}(r_*)\tilde{R}^+_{\ell m\omega}(r_*)\ ,
\label{scalar_sol}
\end{equation}
where 
\begin{align}
    Z^{+}_{\ell m\omega}(r_*)&=\frac{1}{W_{\ell m\omega}}\int^{r_*}_{-\infty}
    \tilde{R}^-_{\ell m\omega}(r'_\star)\tilde{\tau}_{\ell m\omega}(r'_\star)\,\dd r'\star,\label{Z_infty}\\
    Z^{-}_{\ell m\omega}(r_*)&=\frac{1}{W_{\ell m\omega}}\int^{+\infty}_{r_*}
    \tilde{R}^+_{\ell m\omega}(r'_\star)\tilde{\tau}_{\ell m\omega}(r'_\star)\, \dd r'_\star\label{Z_H}\ ,
\end{align}
and $W_{\ell m\omega}= \tilde{R}_{\ell m\omega}^-\tilde{R}^+_{\ell m\omega,r_*}-\tilde{R}^-_{\ell m\omega,r_*}\tilde{R}^+_{\ell m\omega}$.

As in the Schwarzschild case, ${\tilde R}^\pm_{\ell m\omega}$ are two independent solutions of the homogeneous equation associated to Eq.\,\eqref{eq:X}, with boundary conditions (assuming $\omega<\mu_s)$: 
\begin{align}
&\tilde{R}_{\ell m\omega}^-\sim
\begin{cases}
    e^{-i p_\omega r_*}\quad &r\rightarrow r_h\\
    A_{\text{out}} e^{-\kappa_\omega r_*}+A_{\text{in}} e^{\kappa_\omega r_*} \quad & r\rightarrow \infty
\end{cases}\ ,\\[2mm]
&\tilde{R}_{\ell m\omega}^+\sim
\begin{cases}
    B_{\text{in}} e^{-i p_\omega r_*}+B_{\text{out}} e^{i p_\omega r_*} \quad &r\rightarrow r_h\\
    e^{-\kappa_\omega r_*} \quad &r\rightarrow \infty
\end{cases}\,,
\label{eq:X_boundaries}
\end{align}
where $\kappa_\omega=\sqrt{\omega^2-\mu_{\rm s}^2}$, $p_{\omega}=\omega-m a/2Mr_h$ and $r_h=M+\sqrt{M^2-a^2}$ is the radial coordinate at the event horizon.

For equatorial, circular orbits, the source term is
\begin{align}
\tilde{\tau}_{\ell m\omega}=&-\frac{4\pi \hat{d}_0 m_{\rm p}}{\gamma(r_0,\pi/2)}S^*_{\ell m}(\pi/2,0)\frac{\Delta(r_0)}{\rho(r_0)^3}\nonumber\\
&\times\delta(r-r_0)\delta(\omega-\bar\omega_m)
\end{align}
where $\gamma(r_0,\pi/2)$ can be found from \eqref{eq:tgamma} and $\bar\omega_m=\omega_0^R/\gamma+m\Omega_\phi$.
Therefore,
\begin{equation}
    Z_{\,\ell m\omega}^{\pm}(r)={\mathcal C}_{\ell m\omega}{ \tilde{R}}^\mp_{\ell m\omega}(r_{0*})\Theta(\pm(r-r_0))\delta(\omega-\bar\omega_m)\, ,
\label{eq_exprZ_Kerr}
\end{equation}
where
\begin{equation}
    {\mathcal C}_{\ell m\omega}=\frac{4\pi \hat{d}_0 m_p}{W \gamma(r_0,\pi/2)} S^*_{\ell m} \left(\frac{\pi}{2}\right)
\end{equation}
For $r\to2M$,
\begin{equation}
    \tilde{R}_{\ell m\omega}(r)\sim \frac{e^{-ip_\omega r_*}}{\rho(r_0)}{\mathcal C}_{\ell m\omega}\tilde{R}^+_{\ell m\omega}(r_{0*})\delta(\omega-\bar\omega_m)
    \label{eq:RhorKerr}
\end{equation}
while for $r\to\infty$
\begin{equation}
    \tilde{R}_{\ell m\omega}(r)\sim \frac{e^{-\kappa_\omega r_*}}{\rho(r_0)}{\mathcal C}_{\ell m\omega} \tilde{R}^-_{\ell m\omega}(r_{0*})\delta(\omega-\bar\omega_m)\,.
    \label{eq:RinfKerr}
\end{equation}
In Kerr background, the energy flux a infinity and at the horizon, Eq.\,\eqref{eq:E_dot_SC}, reads$\dot E_\pm=\mp \int \dd \Omega\   T_{tr} \Delta(r)$, with $T_{tr}$ given in Eq.\,\eqref{eq:Trt_SC}. By replacing Eqns.\,\eqref{eq:kerr_harm_phi}-\eqref{eq:kerr_four_T} we find that Eq.\,\eqref{eq:solEdot}, derived for Schwazrchild background, holds for Kerr backgdound as well. 

By replacing Eq.\,\eqref{scalar_sol}, we find that - as in the Schwarzschild case - since $\bar\omega_m,\mu_{\rm s}\gg M^{-1}$ and $\bar\omega_m<\mu_{\rm s}$, the flux at infinity vanishes, while
\begin{align}
    \frac{\dd E_-}{\dd t}=&\frac{2d_0^2 m_{\rm p}^2\pi}{\gamma(r_0,\pi/2)^2}\frac{r_0^2}{\rho(r)^4}\nonumber\\
    &\times \sum_{\ell m} \bar{\omega}_m p_{\bar{\omega}_m} {\mathcal R}_{\ell m}(r_0)
    S^2_{\ell m}(\pi/2) \,,\label{eq:fluxKerr}
\end{align}
with ${\mathcal R}_{\ell m}(r_0)$ given in Eq.\,\eqref{eq:ratio_R/W}.
%
%%%%%%%%%%%%%%%%%%%%%%%%%%%%%%%%%%%%%%%%%%%%%%%%%%%%%%%%%%%%%%%%%%%%%%%%%
\section{Solving the wave equation in the high-frequency limit}\label{app:solvingwave}
%%%%%%%%%%%%%%%%%%%%%%%%%%%%%%%%%%%%%%%%%%%%%%%%%%%%%%%%%%%%%%%%%%%%%%%%%
\begin{figure}
    \centering
    \includegraphics[width=0.7\linewidth]{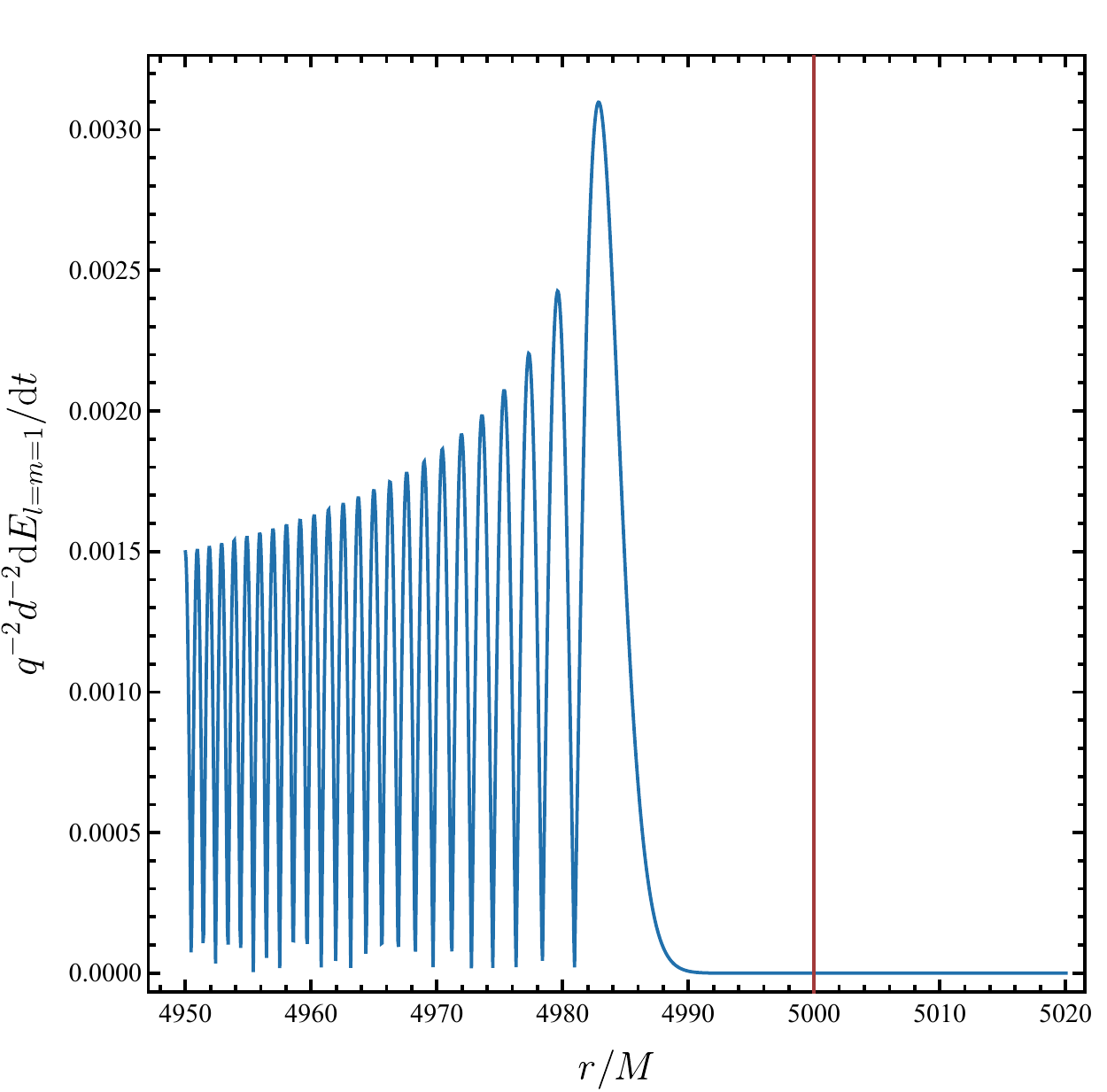}
    \includegraphics[width=0.7\linewidth]{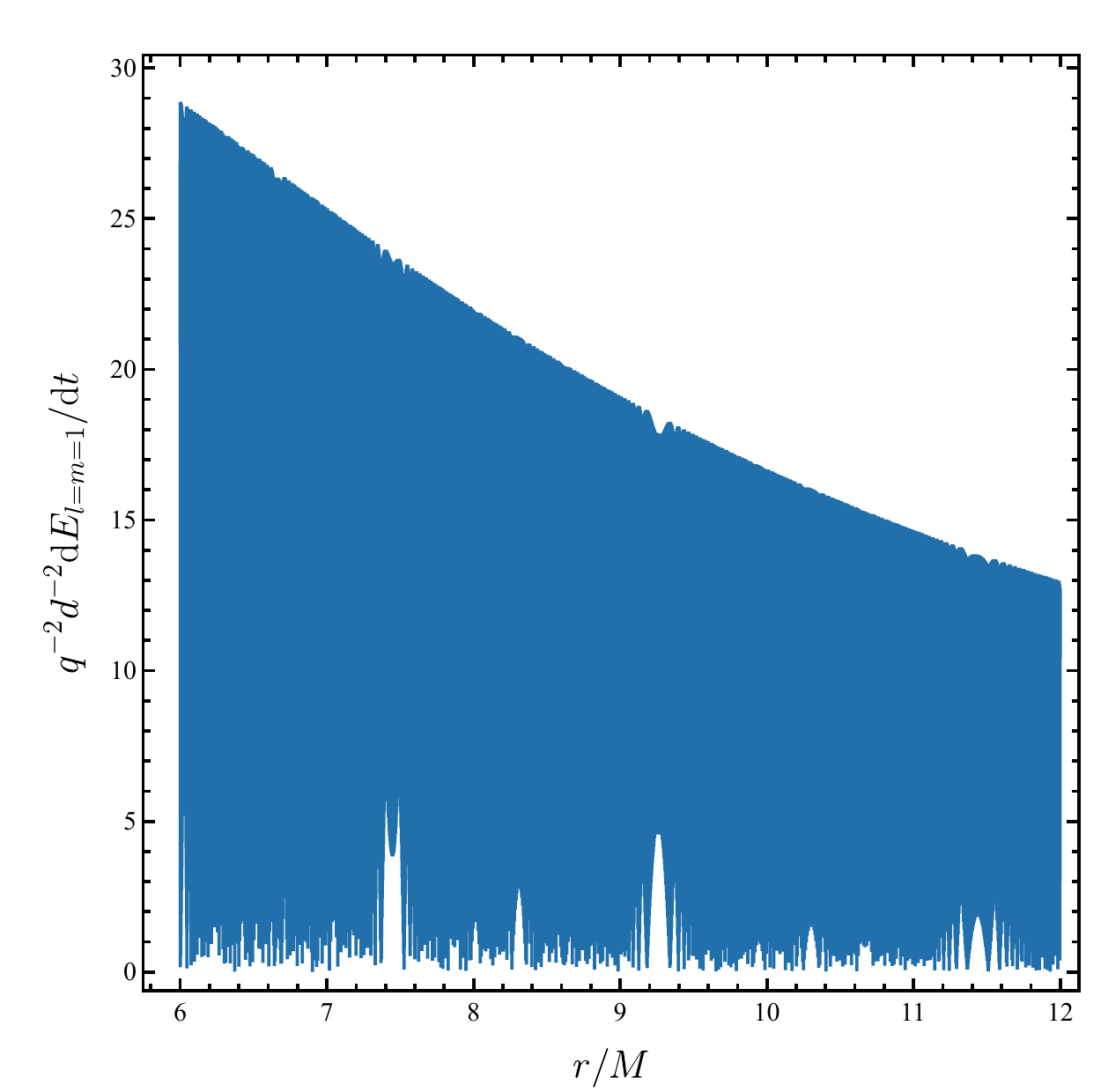}
    \caption{Scalar field energy flux for $\mu_{\rm s} M=2000.0$ and $\omega M=1999.6$. The upper panel shows the flux, normalized with mass ratio and scalar charge, emitted near the zero $r=\bar r$ of the effective potential (vertical line). The lower panel shows the flux for $r\sim(6-12)\,M$, corresponding to the LISA sensitivity band.}\label{fig:flux}
\end{figure}
%%%%%%%%%%%%%%%%%%%%%%%%%%%%%%%%%%%%%%
In this Appendix we discuss how to numerically solve the homogeneous wave equation in the Schwarzschild spacetime\,\eqref{eq:Einstein_Klein_Gordon_Fourier},in the limit of large frequency $\omega M\gg1$ and large scalar mass $\mu_{\rm s}M\gg1$, with $\omega<\mu_{\rm s}$, such that the it reduces to Eq.~\eqref{eq:homlargew}
\begin{equation}
\left[\frac{\partial^2 }{\partial r^2_*}+{\mathcal V}_\omega(r)\right]{\tilde R}^\pm_{\omega}=0 .
\label{eq:RWhigh}
\end{equation}
with the effective potential\,\eqref{eq:calV}, ${\mathcal V}_\omega(r)\simeq\omega^2-(1-2M/r)\mu_{\rm s}^2$.
%----------------------------------------------
\subsection{Approximated potential}
%----------------------------------------------
The presence of two scales ($r\sim M$ and $\omega^{-1}$, $\mu_s^{-1}\ll M$) makes it challenging to numerically solve Eq.~\eqref{eq:RWhigh}. To gain qualitative insight into the solution, we explore two approximations of ${\mathcal V}_\omega(r)$, offering different levels of accuracy in describing the effective potential.

Let's first consider a toy model in which the potential is represented by a step function
\begin{equation}
    \mathcal{V}^{\rm toy}_\omega(r)=\begin{cases}
        \omega^2  &  r\le  \bar r\\
       \omega^2- \mu_{\rm s}^2 & r\ge\bar r
    \end{cases}\ .
\end{equation}
with $\omega M,\mu_{\rm s}M\gg 1$ and $\omega<\mu_{\rm s}$.
In this case, a simple calculation shows that the function $\mathcal{R}$ (see Eq.~\eqref{eq:ratio_R/W}) is characterised is exponentially suppressed, $\mathcal{R}\sim e^{-2\kappa_\omega (r_\star(r_0)-r_\star(\bar r))}$, for $r>r_0$. This leads to a suppression of the emitted flux for quasi-circular orbits with $r_0>\bar r$, while the emission  ``turns on'' when $r_0<\bar r$.

We now consider a better approximation of Eq.\,\eqref{eq:calV}: the analytic potential
\begin{equation}
    {\mathcal V}^{\rm an}_{\omega}=
    \begin{cases}
       \omega^2- \frac{\mu_s^2}{2}\left(1+{e^{-2}}\right) \left[1+e^{\frac{r_*}{2M}} \right]  & r_*<4M
        \\[2mm]  
       \omega^2- \mu_s^2 \left(1-\frac{2M}{r_*}\right) & r_*>4M\ .
    \end{cases}
    \label{eq:V_approximated}
\end{equation}
For the values of $\omega$ and $\mu_{\rm s}$ conisdered, this potential approximates Eq.\,\eqref{eq:calV} within percent for $r$ comparable with $\bar r$. At smaller values of $r$ the the relative discrepancy increases, becoming $\sim 10\%$ for $r_*\sim 4M$, and reaching at most $\sim 30\%$\,\footnote{More accurate approximations of the potential (with discrepancies always withinn few percent) lead to the same qualitative behaviour of the emitted flux, even though it is numerically difficult to perform a computation such as that shown in Fig.\,\ref{fig:flux}.}.

Solving Eq.\,\eqref{eq:RWhigh} with this potential and  the boundary conditions\,\eqref{eq:asymptoci_H} and 
\eqref{eq:asymptoci_inf}, we find that for 
$r_*>4M$,
\begin{equation}
    \tilde{R}^-= e^{-i\omega r_*}{}_2F_1\left(a,-a,1-4Mi\omega,e^{-\frac{r_*}{2M}} \right)\,,
\end{equation}
where $a\equiv -2Mi\omega+\sqrt{2(1+e^{-2})\mu_s^2-4\omega^2}$ and ${}_2F_1(a,b,c,x)$ is the hypergeometric function, while for $r_*<4M$ %
\begin{equation}
\tilde{R}^+=(2\kappa)^{-{M\mu^2}{\kappa}/}(2\kappa r_*) e^{-\kappa r_*} U\left(1-{M\mu^2}/{\kappa},2,2\kappa r_*\right)\,,
\end{equation}
where $U(a,b,x)$ is the confluent hypergeometric function of the second kind. From these expressions we can evaluate ${\mathcal R}(r_0)$ in Eq.\,\eqref{eq:ratio_R/W} and then the energy flux $\dot E_-$ at the horizon of the primary, Eq.\,\eqref{eq:Em_dot}. 

We find the same qualitative behaviour obtained with the toy model:  if there exists a value $r=\bar r$ such that ${\mathcal V}_\omega(\bar r)=0$, the flux is vanishing for $r>\bar r$ (for which ${\mathcal V}_\omega<0$), while there is a large emission for $r<\bar r$ (for which ${\mathcal V}_\omega>0$). If, instead, ${\cal V}_\omega$ is always negative, the flux is always suppressed.

In Fig.\,\ref{fig:flux} we show the ingoing scalar energy flux normalized to mass ratio and scalar charge, as a function of $r$, computed  approximating the effective potential with Eq.\,\eqref{eq:V_approximated}. We set $\mu_{\rm s} M=2000.0$ and $\omega M=1999.6$, corresponding to $\mu_{\rm s}m_{\rm p}=0.02$ and $\omega=\omega_0^R$; for these values, the effective potential has a zero at $\bar r\simeq5\times 10^3$. We find that even for a scalar charge $d_0\gtrsim 10^{-6}$, the scalar flux emitted when $r_0<\bar r$ is orders of magnitudes larger than the gravitational wave flux. We also note that it rapidly increases for smaller values of $r$. 

We remark that the flux shown in Fig.\,\ref{fig:flux} is computed for a given value of the frequency. However, as discussed in Sec.\,\ref{subsec:pert}, the emission in an actual EMRI occurs at a frequency $\bar\omega_m=\omega^R/\gamma+m\Omega_\phi$ which decreases during the inspiral.
%
%----------------------------------------------
\subsection{Does the effective potential change sign in actual EMRIs?}\label{app:proof_neg}
%----------------------------------------------

As discussed in Sec.\,\ref{subsec:flux}, in the case of quasi-circular orbits it turns out that ${\mathcal V}_\omega$ is always negative, and thus the scalar emission never activates. Indeed, for a circular orbit at $r=r_0$ the scalar emission is at frequency $\omega=\bar\omega_m=\omega_0^R/\gamma+m\Omega_\phi$. The contribution to the orbital motion can be neglected, since $|m\Omega_\phi|<\mu_{\rm s}-\omega_0^R<\mu_{\rm s}-\omega_0^R/\gamma$ for $m\lesssim 400$ (and the emission for large harmonic indexes is strongly suppressed, see e.g.\,\cite{Fujita:2005kng,Drasco:2005kz}). The condition for a change in sign in the effective potential, then, reduces to $\gamma^{-2}>(1-2M/r_0)$, which is never satisfied since, for circular orbits in Schwarzschild spacetime, $\gamma^{-2}=1-3M/r_0$. It is worth noting that, if the particle is set at rest at a given time, $\gamma^{-2}=1-2M/r_0$ which prefectly cancels with the redshift of the mass term: in that case, the wave equation behaves as in Minkowski space.

For a general orbit in Schwarzchild spacetime, we still have (we use an overdot to denote derivatives with respect to $t$)
\begin{align}
\gamma^{-2}&=1-\frac{2M}{r_0}-g_{rr}(\dot r)^2-r^2(\dot\theta)^2-r^2\sin^2\theta\Omega_\phi^2\nonumber\\
&<1-\frac{2M}{r_0}\,,\label{eq:disrednorot}
\end{align}
and thus ${\mathcal V}_\omega(r_0)<0$ for all values of $r_0$. 

We can generalize Eq.\,\eqref{eq:disrednorot} to a Kerr background. In this case, the wave equation is given by 
Eqns.\,\eqref{eq:X}-\eqref{eq:potKerr}. For $\omega,\mu_{\rm s}\gg M^{-1}$ we recover Eq.\,\eqref{eq:RWhigh} with effective potential
\begin{equation}
{\mathcal V}_\omega(r)\simeq \omega^2-\frac{2am}{\rho^2}\omega-\frac{\Delta}{\rho^2}\mu_{\rm s}^2\,.
\label{calVKerr}
\end{equation}
To assess the sign of Eq.\,\eqref{calVKerr} for $\omega=\bar\omega_m$, we first note that $|a/\rho^2|<\Omega_\phi$ (Eq.\,\eqref{eq:Kerr:circeqfreq}), and therefore the term $ma/\rho^2$ can be neglected like the term $m\Omega_\phi$. Moreover, we have
\begin{equation}
    \gamma^{-2}=-g_{\mu\nu}{\dot x}^\mu{\dot x}^\nu<-g_{tt}-2g_{t\phi}\Omega_\phi-g_{\phi\phi}\Omega_\phi^2
    \label{eq:maxgamma}
\end{equation}
where we have used the fact that $g_{rr}(\dot r)^2+g_{\theta\theta}(\dot\theta)^2>0$. Since the quantity of the right-hand side of Eq.\,\eqref{eq:maxgamma} is maximum at $\Omega_\phi=-g_{t\phi}/g_{\phi\phi}$, we have
\begin{equation}
\gamma^{-2}<-g_{tt}+\frac{g_{t\phi}}{g_{\phi\phi}}=\frac{\Delta}{\rho^2+\frac{2Mra^2}{\Sigma}\sin^2\theta}_{|r=r_0}<\frac{\Delta}{\rho^2}_{|r=r_0}\,.\label{eq:disredrot}
\end{equation}
Therefore for a general orbit in Kerr spacetime,  ${\mathcal V}_{\bar\omega_m}(r_0)<0$ during the entire EMRI inspiral, in the range of scalar field mass we are considering ($0.001<\mu_{\rm s}m_{\rm p}<0.02$). In the case of circular, equatorial orbits, we have shown that this leads to a suppression of the scalar field energy flux.  
We expect a similar suppression mechanism for 
more complex orbital configurations. 
%%%%%%%%%%%%%%%%%%%%%%%%%%%%%%%%%%%%%%%%%%%%%%%%%%%%%%%%%
%%%%%%%%%%%%%%%%%%%%%%%%%%%%%%%%%%%%%%%%%%%%%%%%%%%%%%%%%
\bibliography{Bibliografia}
\end{document}